\documentclass{aastex701}  

\shorttitle{MapHub}
\shortauthors{Cao, Avestruz, and Wang}
\usepackage{CJKutf8}

\usepackage{graphicx}	
\usepackage{amsmath}	
\usepackage{amssymb}	

\graphicspath{{ratings/}}

\begin{document}

\title{MapHub: Daily Relevance Heatmaps of New arXiv Preprints for a Research Community}

\author[orcid=0000-0002-1699-6944]{Kaili Cao (\begin{CJK*}{UTF8}{gbsn}曹开力\end{CJK*})}
\altaffiliation{Schmidt Fellow}
\affiliation{Michigan Institute for Data and AI in Society (MIDAS), University of Michigan, 500 Church Street, Ann Arbor, MI 48109, USA}
\email[show]{kailicao@umich.edu}

\author[orcid=0000-0001-8868-0810]{Camille Avestruz}
\affiliation{Department of Physics, University of Michigan, 450 Church Street, Ann Arbor, MI 48109, USA}
\affiliation{Leinweber Institute for Theoretical Physics, University of Michigan, 450 Church Street, Ann Arbor, MI 48109, USA}
\email{}

\author[orcid=0000-0002-7633-431X]{Feige Wang}
\affiliation{Department of Astronomy, University of Michigan, 1085 South University Avenue, Ann Arbor, MI 48109, USA}
\email{}


\begin{abstract}
MapHub, the MIDAS arXiv Preprint Hub, assesses each day's new arXiv submissions for every participant in a research community, with a large language model reading each paper against portfolios built for the participants. It publishes the relevance scores, from 0 to 100, as a heatmap that shows whose work each paper touches. In the authors' blind ratings, the top 10\% of papers by MapHub's scores held about 70\% of the raters' own top 5\%, pooled, against about a third for a text-matching baseline. About 6\% of a day's astro-ph papers reach 30 for each participant, and one every three days reaches 60. MapHub runs on a standard Claude subscription and is designed to be useful for any field: \url{https://github.com/kailicao/maphub-public}.
\end{abstract}

\section{Introduction} \label{sec:intro}

Nowadays, each arXiv announcement brings about ninety new astro-ph submissions, and readers cannot attend to all of them equally \citep{2009arXiv0907.4740H}. myADS runs keyword and author queries on new preprints \citep{2007ASPC..377..106H}; recommenders learn a reader's interests from topics, text similarity, or ratings \citep{2019JApA...40...23K, 2025arXiv250408385F}, or adapt inspectable profiles day by day \citep{2026arXiv260607454W}. Open-source projects use large language models to rank each day's listing against one reader's interests, and at Ohio State, agents built from 21 faculty members' publication records each discuss matching daily papers \citep{2026arXiv260329771T}.

These tools serve one reader at a time, or share discussions, rather than publishing how each paper relates to each member. MapHub\footnote{\url{https://kailicao.github.io/maphub-public/}} combines three things: Every new paper is scored for every participant and shown in one heatmap, one row per paper and one column per participant; the profiles are portfolios that participants can check and correct themselves; the scores are public and traceable.

\section{Design} \label{sec:design}

A public prompt can be used to draft each participant's research portfolio from a publication record, and the participant checks it: about 1,500–3,000 words, built from their own papers and plans, listing methods and targets as well as topics. Portfolios stay among participants and the model's provider; scores are public, as participants agree when they join. {\sc Claude Opus 5.5} \citep[at low effort;][]{Anthropic2026} reads each new paper's title, authors, and abstract against the portfolios and returns a relevance score from 100 (``the participant wrote the paper, or would'') through 60 (``would cite it'') and 30 (``it may broaden their view'') to 0, a fit to the portfolio rather than a judgment of quality. MapHub assesses new submissions to its listings (here astro-ph and cs.AI; the listings can be configured).

Since most cells score low, a first pass scores chunks of about 50 papers for all participants at once against short outlines of the portfolios, and only papers scoring 10 or more are rescored against a full portfolio. On two test listings (748 papers), this cost about 36\% of a single full pass and missed none of the 56 cells that pass scored 40 or more. The prompts, code, and tables are public, scores are labeled as AI-generated, and each table's commit records what produced it, so the output can be checked.

\begin{figure*}
    \plotone{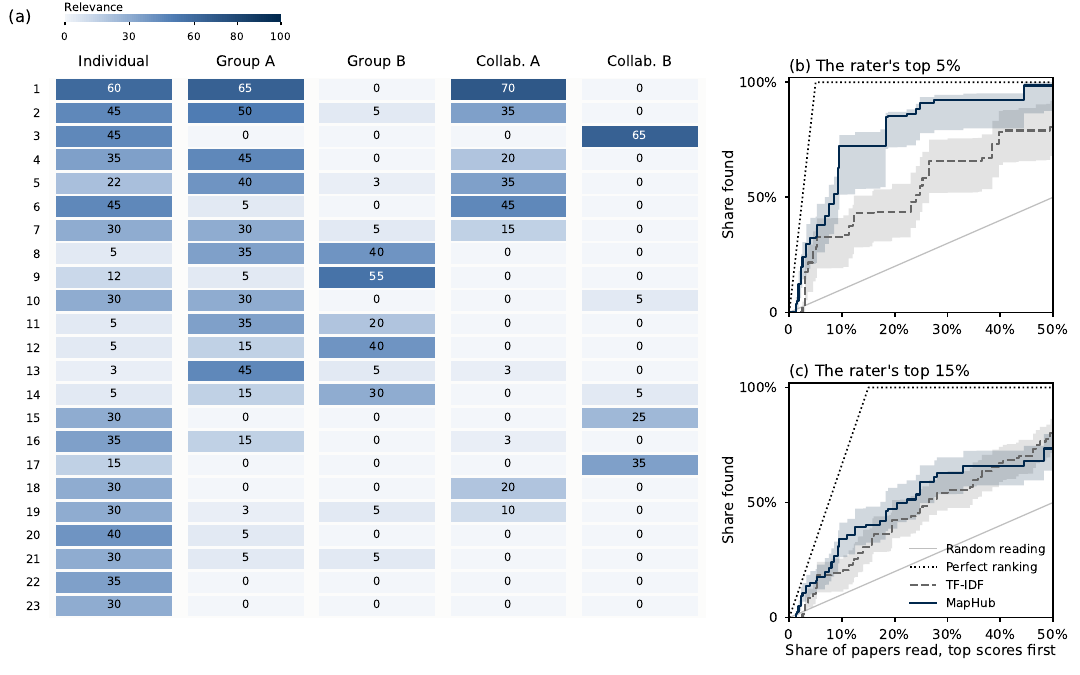}
    \caption{(a) An example day of MapHub, as its viewer shows it: the 23 papers (of 262) that reached 30 for some participant on 2026 August 25, numbered by average score. (b), (c) The share of each rater's top 5\% (b) and top 15\% (c) found, against the share of astro-ph papers read in order of score. Columns are pooled and weighted back to uniform sampling; bands are 68\% bootstrap intervals.
    \label{fig:human}}
\end{figure*}

\section{First results} \label{sec:results}

Seventy listings scored with version 0.8's portfolios and prompt held 11,932 new submissions, scored for one individual, two groups, and two collaborations. Of a day's 87 astro-ph papers, $6.3\% \pm 3.5\%$ (mean $\pm$ standard deviation) reach 30 for each participant and 0.4\% reach 60. Even the individual column, spanning multiple subfields, sees only $7 \pm 3$ astro-ph papers a day at 30 or more: minutes of reading abstracts. And 8\% of astro-ph papers reach 30 for two or more participants (Figure~\ref{fig:human}a). Of a day's 84 cs.AI papers, about two reach 30 for the individual, and almost none for the others.

To test whether the scores save time, the authors rated one shared list blind for themselves or their groups on MapHub's rubric, each interpreting its descriptions in their own way.\footnote{One rater recast its descriptions as a scale from 1 to 5. Only each rater's order is used.} The list held 45 astro-ph papers drawn with probabilities rising with a paper's highest score ($\propto 2^{s/20}$), five of them shown twice, and 10 cs.AI papers spanning the individual column's scores. Papers by the participants were left out; the portfolios were fixed on 2026 October 14 and cite none of the rated papers. Each rating is weighted by the inverse of its paper's chance of being drawn, so the results estimate what reading these listings would find. A paper counts as of interest if it is in the rater's top 5\% (or 15\%) of the population by their own ratings.

Reading in MapHub's order finds the rater's top 5\% quickly (Figure~\ref{fig:human}b): MapHub's top 5\% of papers hold about a third of it, and its top 10\% about 70\%, against 33\% for a term-frequency baseline (TF-IDF\footnote{\url{https://scikit-learn.org/stable/modules/generated/sklearn.feature_extraction.text.TfidfVectorizer.html}} of title and abstract against the portfolio) reading the same share. MapHub's lead is 39 points at 10\% read (95\% interval: 11--60; ahead in every bootstrap draw). The top 15\% is spread more widely: MapHub's top 10\% holds about a third of it, against about 20\% for TF-IDF, and the two converge when about 40\% of the listing is read (Figure~\ref{fig:human}c). MapHub's cutoff of 30 means reading 4\%--9\%, and over 90\% of the papers above it were in the rater's top 15\%.

\section{Discussion} \label{sec:disc}

A department's heatmap could help an Astro Coffee-style discussion\footnote{\url{https://astronomy.osu.edu/coffee}} of the whole listing: Papers touching several columns are natural topics, and their highest-scoring participants natural people to ask. Portfolios drawn from publication records may miss interests not yet published: The groups' portfolios name machine learning only as a tool for their science, so MapHub may miss AI papers their members want. Adding these interests to the portfolios would fix this; meanwhile, the individual's column flags AI papers. Scores stay comparable across days, since every paper is read against the portfolio on one rubric, and rescoring changed them by 1.7 points on average.

MapHub is still under development, and the test is small. Its ratings are subjective: The first author, who built the tool, rated three of the five columns, and a rating shifts with the papers read just before it. Each rater's top 5\% holds few papers, and defining interest by percentile assumes every rater has some.

\begin{acknowledgments}
K.C. was supported by the Eric and Wendy Schmidt AI in Science Postdoctoral Fellowship, a program of Schmidt Sciences. C.A. acknowledges support from the Leinweber Institute for Theoretical Physics at the University of Michigan and DOE grant DE-SC009193. This work was, in part, enabled by the Claude Team Plan for Scientists.
\end{acknowledgments}

\section*{AI Disclosure}

Claude (Anthropic) was used to draft parts of this note's text. Responsibility for the final text lies entirely with the authors.

\section*{Data Availability}

MapHub's code, documentation, and tables are on GitHub. This note describes version 0.8 \citep{10.5281/zenodo.23539894}.

\bibliography{main}{}

@ARTICLE{2026arXiv260607454W,
       author = {{Wang}, Fuqiang and {Tan}, Song and {Guo}, Zheng and {Fu}, Jiaohao and {Xu}, Xinglong and {Yu}, Bihui and {Dong}, Jie and {Sun}, Zheng and {Li}, Siyuan and {Wei}, Jingxuan and {Tan}, Cheng},
        title = "{PaperFlow: Profiling, Recommending, and Adapting Across Daily Paper Streams}",
      journal = {arXiv e-prints},
         year = 2026,
        month = jun,
          eid = {arXiv:2606.07454},
        pages = {arXiv:2606.07454},
          doi = {10.48550/arXiv.2606.07454},
archivePrefix = {arXiv},
       eprint = {2606.07454},
 primaryClass = {cs.IR},
       adsurl = {https://ui.adsabs.harvard.edu/abs/2026arXiv260607454W}
}

@ARTICLE{2025arXiv250408385F,
       author = {{Flicke}, Markus and {Angrabeit}, Glenn and {Iyengar}, Madhav and {Protsenko}, Vitalii and {Shakun}, Illia and {Cicvaric}, Jovan and {Kargi}, Bora and {He}, Haoyu and {Schuler}, Lukas and {Scholz}, Lewin and {Agnihotri}, Kavyanjali and {Cao}, Yong and {Geiger}, Andreas},
        title = "{Scholar Inbox: Personalized Paper Recommendations for Scientists}",
      journal = {arXiv e-prints},
         year = 2025,
        month = apr,
          eid = {arXiv:2504.08385},
        pages = {arXiv:2504.08385},
          doi = {10.48550/arXiv.2504.08385},
archivePrefix = {arXiv},
       eprint = {2504.08385},
 primaryClass = {cs.CL},
       adsurl = {https://ui.adsabs.harvard.edu/abs/2025arXiv250408385F}
}

@ARTICLE{2019JApA...40...23K,
       author = {{Kerzendorf}, Wolfgang E.},
        title = "{Knowledge discovery through text-based similarity searches for astronomy literature}",
      journal = {Journal of Astrophysics and Astronomy},
         year = 2019,
        month = jun,
       volume = {40},
       number = {3},
          eid = {23},
        pages = {23},
          doi = {10.1007/s12036-019-9590-5},
archivePrefix = {arXiv},
       eprint = {1705.05840},
 primaryClass = {cs.DL},
       adsurl = {https://ui.adsabs.harvard.edu/abs/2019JApA...40...23K}
}

@ARTICLE{2009arXiv0907.4740H,
       author = {{Haque}, Asif-ul and {Ginsparg}, Paul},
        title = "{Positional effects on citation and readership in arXiv}",
      journal = {Journal of the American Society for Information Science and Technology},
         year = 2009,
       volume = {60},
        pages = {2203-2218},
          doi = {10.1002/asi.21166},
archivePrefix = {arXiv},
       eprint = {0907.4740},
}

@INPROCEEDINGS{2007ASPC..377..106H,
       author = {{Henneken}, E. and {Kurtz}, M.~J. and {Eichhorn}, G. and {Accomazzi}, A. and {Grant}, C.~S. and {Thompson}, D. and {Bohlen}, E. and {Murray}, S.~S.},
        title = "{myADS-arXiv -- a Tailor-made, Open Access, Virtual Journal}",
    booktitle = {Library and Information Services in Astronomy V},
         year = 2007,
       editor = {{Ricketts}, S. and {Birdie}, C. and {Isaksson}, E.},
       series = {Astronomical Society of the Pacific Conference Series},
       volume = {377},
        month = oct,
    publisher = {ASP},
        pages = {106},
          doi = {10.48550/arXiv.cs/0608027},
archivePrefix = {arXiv},
       eprint = {cs/0608027},
 primaryClass = {cs.DL},
       adsurl = {https://ui.adsabs.harvard.edu/abs/2007ASPC..377..106H}
}

@MISC{Anthropic2026,
       author = {{Anthropic}},
        title = "{System Card: Claude Opus 5.5}",
         year = 2026,
        month = sep,
          url = {https://www.anthropic.com/claude-opus-5-5},
         note = {System card published 2026 September 22},
}

@ARTICLE{2026arXiv260329771T,
       author = {{Ting}, Yuan-Sen},
        title = "{Your Outie Is a Wonderful Astronomer: Macrodata Refinement of the Astro-ph ArXiv Feed at Phermon Industries}",
      journal = {arXiv e-prints},
         year = 2026,
        month = mar,
          eid = {arXiv:2603.29771},
        pages = {arXiv:2603.29771},
          doi = {10.48550/arXiv.2603.29771},
archivePrefix = {arXiv},
       eprint = {2603.29771},
 primaryClass = {astro-ph.IM},
       adsurl = {https://ui.adsabs.harvard.edu/abs/2026arXiv260329771T}
}

@misc{10.5281/zenodo.23539894,
  doi = {10.5281/ZENODO.23539894},
  url = {https://zenodo.org/doi/10.5281/zenodo.23539894},
  author = {{Cao}, Kaili},
  title = {kailicao/maphub-public: MapHub: code and tables of the research note's rating study (October 2026)},
  publisher = {Zenodo},
  year = {2026},
  copyright = {Creative Commons Attribution 4.0 International}
}
\bibliographystyle{aasjournalv7}

\end{document}